\documentclass[a4paper]{jpconf}
\usepackage{graphicx}
\begin{document}
\title{Hybrid quantum systems of atoms and ions}

\author{Christoph Zipkes, Lothar Ratschbacher, Stefan Palzer, Carlo Sias, Michael K{\"o}hl}

\address{Cavendish Laboratory, University of Cambridge, JJ Thomson Avenue, Cambridge CB3 0HE, United Kingdom}


\begin{abstract}
In recent years, ultracold atoms have emerged as an exceptionally controllable experimental system to investigate fundamental physics, ranging from quantum information science to simulations of condensed matter models. Here we go one step further and explore how cold atoms can be combined with other quantum systems to create new quantum hybrids with tailored properties. Coupling atomic quantum many-body states to an independently controllable single-particle gives access to a wealth of novel physics and to completely new detection and manipulation techniques. We report on recent experiments in which we have for the first time deterministically placed a single ion into an atomic Bose Einstein condensate. A trapped ion, which currently constitutes the most pristine single particle quantum system, can be observed and manipulated at the single particle level. In this single-particle/many-body composite quantum system we show sympathetic cooling of the ion and observe chemical reactions of single particles in situ.
\end{abstract}

\section{Introduction}
Hybrid systems allow the exploration of physics far beyond that which can be studied in the individual components. This has been highlighted, for example, by the immersion of distinguishable particles into a quantum liquid which has contributed significantly to our understanding of many-body systems. In liquid helium vortex lattices have been observed using charged particles as markers for the vortex lines \cite{Yarmchuk1979}. For future investigations with distinct single particles in quantum matter the degree of control achievable over these particles will play a decisive role. To this end, the excellent control over the external and internal degrees of freedom of a single ion trapped in a Paul trap appears highly promising. Immersed into a quantum gas a single trapped ion not just acts as a probe but could be used for local manipulation. Numerous phenomena have been foreseen, including sympathetic cooling \cite{Makarov2003}, the nucleation of localized density fluctuations in a Bose gas \cite{Cote2002,Massignan2005,Goold2009}, scanning probe microscopy with previously unattainable spatial resolution \cite{Kollath2007,Sherkunov2009}, and hybrid atom-ion quantum processors \cite{Idziaszek2007}.

\section{Combining atom trapping and ion trapping}
Ion trapping and atom trapping, even though both originating from the desire to control atoms with great experimental precision, have developed very much independently over the past 30 years. Even though there are some common grounds in experimental techniques, such as laser cooling, only little cross-pollination has been seen. Trapped ions, on one hand, have had a tremendous impact on quantum information processing and precision spectroscopy as the best controllable few particle systems. On the other hand, cold atoms and in particular quantum degenerate gases have profoundly stimulated research on quantum matter at ultralow temperatures. The experimental combination of ion trapping and atomic quantum gases has been demonstrated only very recently \cite{Zipkes2010,Schmid2010}.

\begin{figure}[h]
\includegraphics[width=16pc]{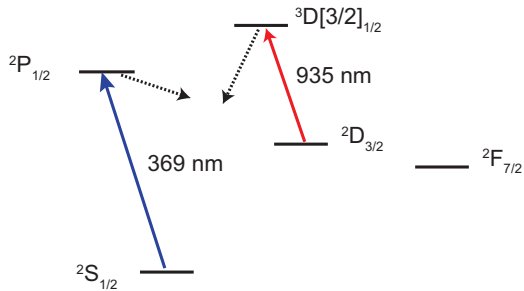}\hspace{2pc}%
\begin{minipage}[b]{14pc}\caption{\label{Yb}Level structure of singly ionized Ytterbium with the relevant energy levels shown.}
\end{minipage}
\end{figure}

In our experimental setup, the ion is confined in the ponderomotive potential (pseudo-potential) of a linear Paul trap which is created from a very rapidly oscillating electric quadrupole field. The quadrupole electric field is generated by four parallel electrodes oriented along the x-direction is $\vec{E}(x,y,z)=\frac{V_{RF}}{R^2}(0,y,-z)\cos(\Omega t)$. The trap operates at an RF frequency of $\Omega=2\pi\times 42.7$\,MHz, well above the typical frequencies used in evaporative cooling of the neutral atoms. The distance between the ion and the RF electrodes is $R=0.4$\,mm and the applied voltage of $V_{RF}\approx 170$\,V gives rise to a radial confinement of $\omega_{\perp,ion}= 2\pi\times 1.5\cdot 10^5$\,Hz. The axial confinement $\omega_{x,ion}= 2\pi\times 4\cdot 10^4$\,Hz is provided by two endcap electrodes charged to a few volts. A single $^{174}$Yb$^+$ ion is loaded into the ion trap by isotope-selective Doppler-free two-photon ionization from a neutral ytterbium atomic beam using resonant light at 398\,nm and 369.5\,nm \cite{Balzer2007}. We pulse the atomic oven generating the atomic beam for 100\,ms only which prevents deteriorating effects of the ion loading procedure on the ultrahigh vacuum ($<2\times 10^{-11}$\,mbar) in our experimental setup. Subsequently, the ion is laser cooled on the $^2S_{1/2}\rightarrow\, ^{2}P_{1/2}$ transition at 369.5\,nm (see Figure \ref{Yb}). Repumping light is provided at 935\,nm to clear population out of the metastable $^2D_{3/2}$ state. From time to time the ion may fall into the dark $F_{7/2}$ state in which case we discard the ion and load a new one. We continually cancel residual dc electric fields, which would result in excess micromotion of the ion, to better than 0.5\,V/m using a photon-correlation fluorescence technique \cite{Berkeland1998}.

\begin{figure}[h]
\hspace{2pc}\includegraphics[width=20pc]{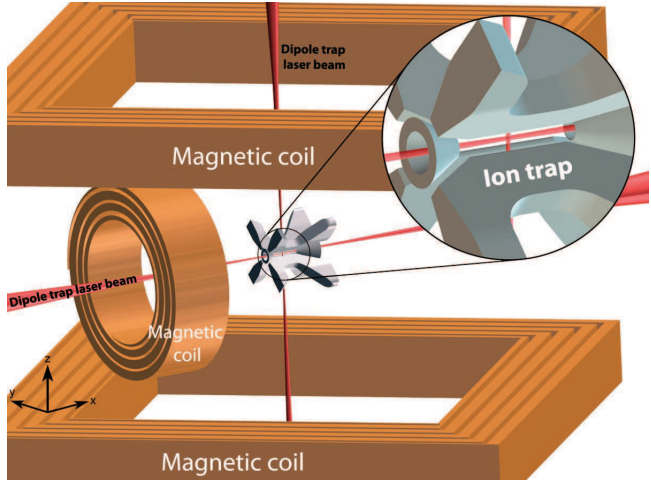}\hspace{2pc}%
\begin{minipage}[b]{14pc}\caption{\label{FigureOverview}Drawing of the apparatus. Shown are the solenoids providing magnetic confinement to the atoms and the linear Paul trap sandwiched in between. The drawing is to scale and the spacing between the magnetic field coils is 36\,mm. The inset shows the ion trap electrodes in detail and the laser beams. Adapted from reference \cite{Zipkes2010}.}
\end{minipage}
\end{figure}

Ultracold $^{87}$Rb atoms are prepared using standard techniques of laser-cooling and evaporative cooling in a magnetic trap \cite{Palzer2009}, approximately 8\,mm away from the ion's position (see Figure \ref{FigureOverview}). After reaching a temperature of $\sim 1\,\mu$K, which is slightly above the critical temperature for Bose-Einstein condensation, we transport the neutral atomic cloud from its initial position into the ion trap by displacing the potential minimum of the magnetic Ioffe trap using suitably timed changes of the electrical currents in the solenoids. The neutral atoms enter the ion trap through a 700\,$\mu$m wide bore in the end cap electrode, moving along the symmetry axis of the linear ion trap. The final magnetic trap has trap frequencies of $\omega_x=2 \pi \times (7.8\pm 0.1)$\,Hz, $\omega_y=2 \pi \times (29.3\pm 1.8)$\,Hz, and $\omega_z=2 \pi \times (29.7\pm0.2)$\,Hz.

\begin{figure}[h]
\includegraphics[width=20pc]{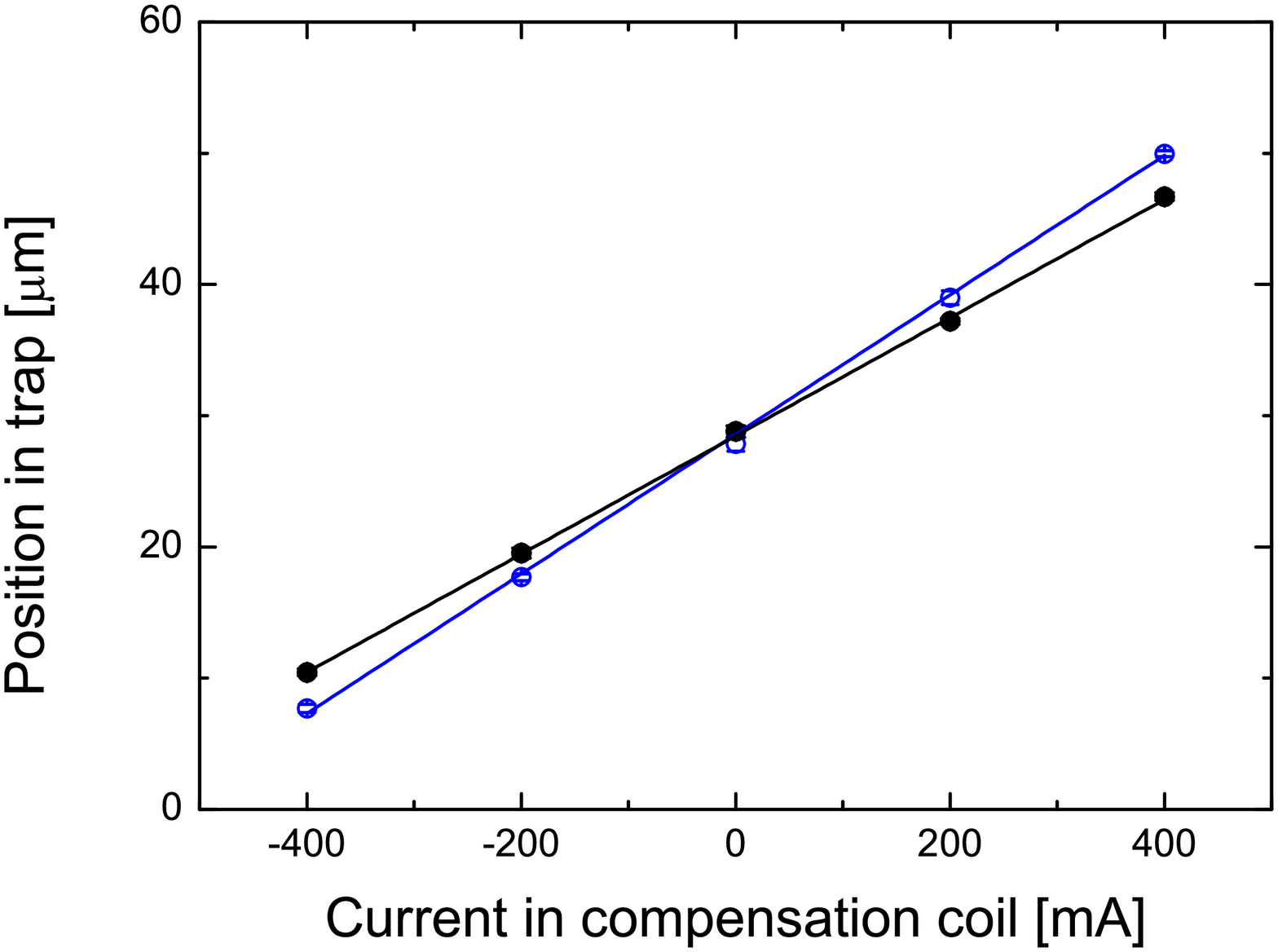}%
\includegraphics[width=20pc]{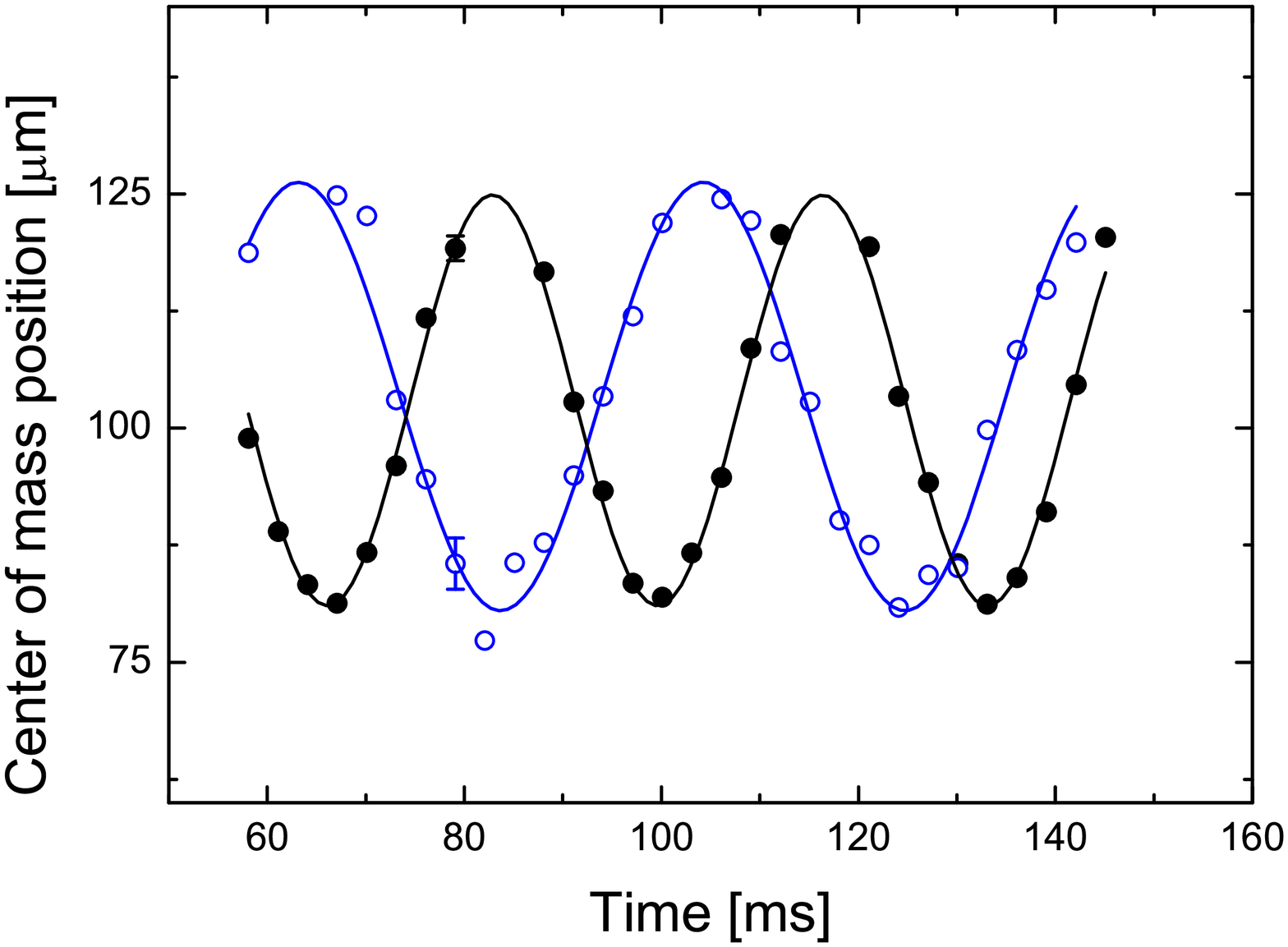}%
\caption{\label{FigOscillation} Left: Shift of the atom cloud in situ upon applying a homogeneous magnetic offset field. The filled symbols show the reference measurement in the absence of the ion trap potential. When the ion trap is switched on (open symbols), we observe that the zero position does not shift but only the slope of the displacement changes. This implies that at the equilibrium position the ion trap does not impose an additional force on the atoms and that the traps are well aligned. Right: Dipole oscillations of neutral atoms in the magnetic trap in the absence (filled symbols) and the presence (open symbols) of the ion trapping potential. Data are taken after 25\,ms of ballistic expansion.}
\end{figure}

The electric field $\vec{E}(\vec{r})$ of the ion trap provides also a potential for the neutral atoms via the quadratic dc-Stark effect (or polarisation potential): $V_{dc}(x,y,z)=-\frac{\alpha}{2}\langle |\vec{E}(\vec{r})|^2 \rangle = -\frac{\alpha V_{RF}^2}{4R^4}(y^2+z^2)$. Here $\alpha$ denotes the dc-polarizability of the neutral atoms and  $\langle . \rangle$ denotes the time average over one period of the ion trap drive frequency. We verify that the RF-potential of the ion trap is centred in the magnetic trap of the neutral atoms by measuring the electric field induced displacement of the neutral atom cloud in-situ. Since the dc-Stark potential is harmonic, the force on the atoms is linear in position and zero when the atoms are located exactly at the origin of the electric quadrupole. We calibrate the measurement by applying a homogeneous offset magnetic field and monitor the shift of the position of the atoms in situ without the ion trap  switched on (see Figure \ref{FigOscillation}). Repeating the same measurement with the ion trap switched on reveals the same  zero-position but a different slope due to the dc-Stark effect indicating that the trap centre positions are well matched.

The dc-Stark effect gives rise to a weak anti-trapping harmonic potential. We have measured the anti-confining effect by observing a frequency shift of the collective dipole mode of the neutral atoms in the magnetic trap in the presence of the electric field (see Figure \ref{FigOscillation}). The measured trapping frequency with the ion trap on is $\tilde{\omega}_z=2\pi \times (24.4\pm0.2)$\,Hz. This results in an effective anti-trapping frequency of $\omega_M=2\pi \times 17$\,Hz due to the ion trap.

\section{Elastic collisions}
Atom-ion scattering is dominated by the polarization potential, which asymptotically behaves as $V(r)=-\frac{C_4}{2 r^4}$, and a hard-core repulsion at the length scale of the Bohr radius. Here $r$ is the internuclear separation, $C_4=\frac{\alpha q^2}{(4\pi\epsilon_0)^2 }$, $q$ denotes the charge of the ion and $\epsilon_0$ is the vacuum permittivity. Atom-ion scattering was theoretically investigated as early as 1905 when Langevin calculated the drift velocity of ions in a buffer gas. The Langevin model is based on classical mechanics and the predicted cross section is $\sigma_{L}=\pi \sqrt{2 C_4/E}$, leading to an energy independent collision rate constant. Langevin-type collisions occur in close encounters between an atom and ion when the impact parameter is below a critical value $b_c=(2 C_4/E)^{1/4}$ \cite{Vogt1954} and exhibit a large momentum transfer between the atom and ion. Going beyond the Langevin model, the quantum mechanical description of scattering yields the angular dependence of the differential cross section in more detail. In the energy range under consideration, elastic scattering from the polarization potential involves several partial waves since the s-wave scattering regime of atom-ion collisions is only at sub-microkelvin temperatures \cite{Cote2000,Idziaszek2009}. The semi-classical approximation of the full quantum mechanical elastic scattering cross section for a collision energy $E$ is $\sigma(E)=\pi(1+\pi^2/16)\left(\frac{\mu C_4^2}{\hbar^2  E}\right)^{1/3}$ \cite{Massey1971,Cote2000} ($\mu$ denotes the reduced mass). This includes in particular a large contribution towards forward scattering under very small angles \cite{Zhang2009}, mainly from the centrifugal barrier.

To study collisions between the ion and atoms in a Bose-Einstein condensate we first produce a Bose-Einstein condensate inside the ion trap. To this end we load the neutral atoms into an optical dipole trap formed by two crossed laser beams operating at 935\,nm wavelength centred onto the ideal position of the ion while the ion is displaced by 140\,$\mu$m along the x-direction. Each of the beams is focused to a waist of approximately 70\,$\mu$m and powered by up to a few hundred milliwatts. In the dipole trap we perform evaporative cooling by lowering the laser intensity to reach Bose-Einstein condensates of $3\times 10^4$ atoms. The final trap frequencies of the optical trap are $\omega_\textrm{x,opt}=2\pi\times 51$\,Hz, $\omega_\textrm{y,opt}=2\pi\times 144$\,Hz, $\omega_\textrm{z,opt}=2\pi\times 135$\,Hz. We apply a homogeneous bias magnetic field of less than 1\,G to prevent depolarization of the neutral atomic sample.

\begin{figure}
\includegraphics[width=20pc]{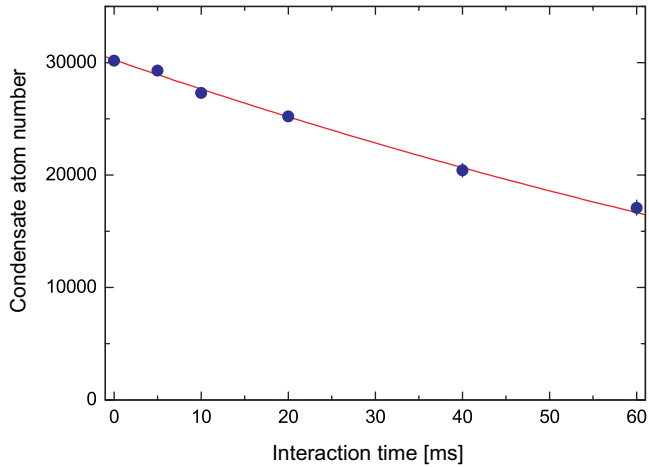}\hspace{2pc}
\begin{minipage}[b]{14pc}\label{Figurecooling}
 \caption{Atom number loss of a Bose-Einstein condensate due to collisions with a single ion. The solid line is a fit using a theoretical model (see \cite{Zipkes2010}) and is used to determine the cross section $\sigma_{al}$ for neutral atom loss. Each data point is averaged over approximately 40 realizations and the statistical error is given. The bare atom loss rate without the ion present has been subtracted. Data taken from reference \cite{Zipkes2010}.}
 \label{fig2}
\end{minipage}
\end{figure}

In order to observe interactions between an ion and a Bose-Einstein condensate we quickly move the pre-cooled ion over a distance of $\Delta x=140\,\mu$m into the center of the Bose-Einstein condensate. The potential depth of the optical dipole trap is 1\,$\mu$K which is less than the vibrational level spacing of the ion trap potential. Therefore, every collision will result in a neutral atom loss. We monitor this loss as a function of the interaction time. The data are displayed in Figure \ref{fig2}. From the decay of the atom number we estimate the cross section for atom loss to be $\sigma_{al}=(2.2\pm0.2)\times10^{-13}$\,m$^{2}$. We have time-resolved the dynamics of the immersion cooling process by immersing a hot ion directly into the Bose-Einstein condensate. For this measurement we partially suppress precooling of the ion in the neutral thermal atom cloud by displacing the ion along the diagonal (\textbf{x+y})-direction. As a result, the initial temperature of the ion is approximately 4\,K. Then we quickly move the ion into the Bose-Einstein condensate, wait for a variable interaction time, release the neutral atoms, and measure the ion's temperature using a fluorescence method \cite{Wesenberg2007,Epstein2007} (see Figure \ref{Figurecooling}). The fluorescence method, even though limited in its temperature resolution, is a reliable and independent method in this temperature range. We observe sympathetic cooling of the ion in a Bose-Einstein condensate on a time scale of a few tens of milliseconds. After 60\,ms the ion has reached a temperature of $T=(0.6\pm0.7)$\,K, corresponding to a temperature as low as our resolution limit.

\section{Charge exchange reactions}

\begin{figure}[htbp]
\includegraphics[width=20pc]{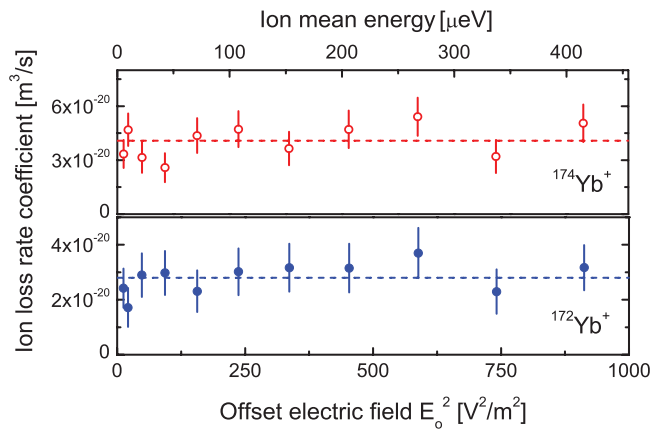} \hspace{2pc}
\begin{minipage}[b]{14pc}\caption{\label{chargeexchange}Measured ion loss probability as a function of offset electric field. Each data point is averaged over approximately 100 repetitions of the experiment. The dashed lines show the average of the data. Data taken from reference \cite{Zipkes2010b}.}
\end{minipage}
\end{figure}

Charge exchange is one of the most fundamental reaction processes in atom-ion scattering. Most recently, with the progress into the regime of cold collisions, theory \cite{Cote2000,Makarov2003,Idziaszek2009,Zhang2009} and experiments \cite{Grier2009,Zipkes2010b} on charge exchange have regained significant attention. In our case, the Rb ($5s\,^2S_{1/2}$) atom in the hyperfine ground state $|F=2,m_F=2 \rangle$ and the Yb$^+$ ($6s\,^2S_{1/2}$) ion in the electronic ground state $|J=1/2,m_J=\pm 1/2\rangle$ collide in the excited A$^1\Sigma^+$ singlet or the a$^3\Sigma^+$ triplet state of the (RbYb)$^+$ molecular potential. The electronic ground state $X^1\Sigma^+$, which asymptotically corresponds to Rb$^+$ ($4p^6\,^1S_0$) and neutral Yb ($6s^2\,^1S_0$), lies 2.08\,eV below. The most striking experimental observation of a charge exchange process is the disappearance of the Yb$^+$ fluorescence which we detect after the interaction with the neutral atoms. We have performed an energy-resolved study of the charge exchange reaction rate by tuning the kinetic energy of the ion \cite{Zipkes2010b}. To this end we apply a homogeneous electric offset field $E_o$ to the ion trap and induce excess micromotion of the ion with an energy $\sim q^2 E_o^2/(m_{ion} \omega_\perp^2)$. We have calibrated this energy using the same laser fluorescence technique that we employed to monitor the ion's temperature during sympathetic cooling. For two different isotopes, $^{172}$Yb and $^{174}$Yb, we find the rate constant to be independent of the collision energy, as predicted \cite{Langevin1905,Vogt1954} (see Figure \ref{chargeexchange}). The average rate constants of $R_{172}=(2.8\pm0.3)\times10^{-20}$\,m$^3$/s and $R_{174}=(4.0\pm0.3)\times10^{-20}$\,m$^3$/s are five orders of magnitude smaller than in the homonuclear case \cite{Grier2009} and of the same order of magnitude as predicted for the similarly asymmetric system (NaCa)$^+$ \cite{Makarov2003}. Because of an additional systematic error between the measurements for the two isotopes of 15$\%$, due to uncertainty of the density determination of the thermal cloud, the difference of the rates for the two different isotopes is statistically not significant.

\begin{figure}[htbp]
\includegraphics[width=.9\columnwidth]{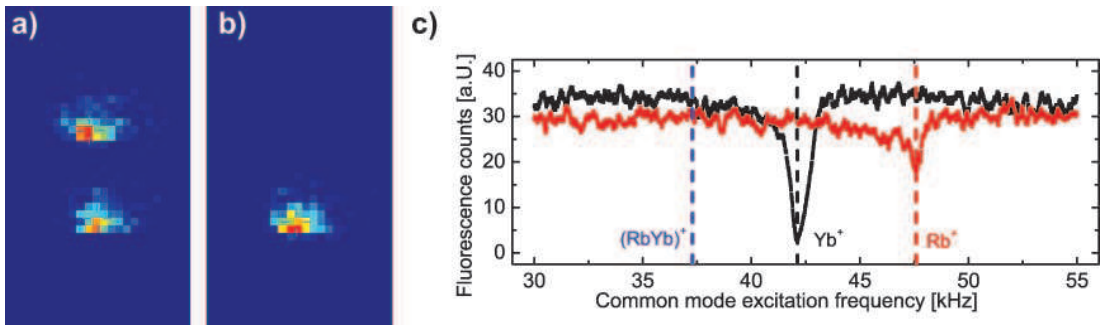}
\caption{\label{massspectrometry} {\bf a)} Ion crystal of two Yb$^+$ ions. {\bf b)} A dark ion has been created. {\bf c)} Mass spectrometry signal together with dashed lines indicating the expected frequencies. The black trace corresponds to two Yb$^+$ ions. For one Rb$^+$ ion and one Yb$^+$ ion (red curve) the collective mode frequency is higher. No signals at the resonance for Yb$^+$ and (RbYb)$^+$ were obtained. Mass spectrometry data taken from reference \cite{Zipkes2010b}.}
\end{figure}

Charge exchange can occur by emission of a photon (radiative charge exchange) or as a nonadiabatic transition between molecular levels. For low temperatures, radiative charge exchange has been predicted to be the dominating process for (NaCa)$^+$ \cite{Makarov2003}. In order to investigate the reaction products of the charge exchange process we perform mass spectrometry. To this end, we load two Yb$^+$ ions into the ion trap and overlap them with the neutral cloud. In the cases in which only one of the two ions undergoes a reaction, the other ion serves to identify the reaction product by measuring a common vibrational mode in the ion trap \cite{Drewsen2004,Willitsch2008}. We excite the axial secular motion of Yb$^+$ by applying intensity modulated light at 370\,nm at an angle of 60$^\circ$ relative to the symmetry axis of the Paul trap and monitor the fluorescence. The frequency of the intensity modulation is swept linearly from 30\,kHz to 55\,kHz in 2.5\,s. When the intensity modulation coincides with a collective resonance of the secular motion, the ions are heated and the fluorescence reduces (see Fig. \ref{massspectrometry}). The collective mode for a Rb$^+$ and a Yb$^+$ ion in the trap is $13\%$ above the bare Yb$^+$ mode at 42\,kHz, whereas the mode of (RbYb)$^+$ and Yb$^+$ would be $12\%$ lower in frequency. In the 486 cases in which one Yb$^+$ ion is lost, we find approximately 30$\%$ probability for the production of cold Rb$^+$ and $70\%$ for a complete loss. If the reaction products take up 2\,eV as kinetic energy in a non-radiative charge exchange they would be lost due to the finite depth ($\sim 150$\,meV) of our ion trap. We have not observed the formation of a charged molecule in this process.

\section{Conclusion}
In conclusion, we have demonstrated the successful realization of a hybrid system of a single trapped ion and a quantum degenerate atomic gas. We have observed elastic collisions between the atoms and the ion leading to sympathetic cooling of the ion in an ultracold neutral environment. Moreover, we have observed charge exchange reactions between a single ion and ultracold atoms and analyzed the reaction products. Our results provide an excellent starting point for future experiments targeting local probing of atomic quantum fluids with a single particle and the full quantum control of chemical reactions at the single particle level.

\ack We thank {EPSRC} (EP/F016379/1, EP/H005676/1), {ERC} (Grant number 240335), and the Herchel Smith Fund ({C.S.}) for support.


\section*{References}
\providecommand{\newblock}{}

\end{document}